\documentstyle[amssymb,revsymb,aps,prb,12pt]{revtex}
\def\beq{\begin{equation}}
\def\eeq{\end{equation}}
\begin{document}
\draft
\title{Dipole and Quadrupole Skyrmions in S=1 (Pseudo)Spin Systems}
\author{N. A. Mikushina and A. S. Moskvin}
\address{Department of Theoretical Physics, Ural State University, 620083, 
Ekaterinburg, Russia}
\date{\today}
\maketitle

\begin{abstract}
In terms of spin coherent states  we have investigated topological defects
 in 2D  $S=1$ (pseudo)spin quantum system
with the bilinear and biquadratic isotropic exchange in the continuum limit.
The proper Hamiltonian of the model can be written as bilinear in
 the generators of SU(3) group (Gell-Mann matrices). The knowledge of such group structure
 allows us to obtain some new exact analytical results.
Analysing the proper classical model we arrive at different skyrmionic
solutions with finite energy and the spatial distribution of spin-dipole and/or
spin-quadrupole moments termed as dipole, quadrupole, and dipole-quadrupole
skyrmions, respectively. Among the latter  we would like note the in-plane
vortices with the in-plane distribution of spin moment, varying spin length,
and the non-trivial distribution of spin-quadrupole moments.
\end{abstract}


%\vspace{0.2cm}

%Keywords: 2D classical and quantum spin systems; topological defects;
% skyrmions

%\vspace{0.2cm}

\newpage {}

\section{Introduction}

Different topological defects  play an important role both in low-energy
 (spin excitations, domain walls, superfluidity/superconductivity)
and high-energy physics from heavy ion collisions to cosmological scenarios
\cite{Thouless,Volovik,Rajantie}. Theoretical approach to its description
traditional for  strongly correlated systems like quantum (pseudo)spin ones,
starts from either (pseudo)spin Hamiltonian with subsequent
  reduction to either classical models with solutions like  in-plane or
out-of-plane vortices, and skyrmions. The latter represent the solutions of
non-linear $\sigma$-model with classical 2D Hamiltonian 
\begin{equation}
 H_{0}= J\int
d^2\vec r\left[\sum_{i=1}^{3}(\vec{\nabla}n_i)^2\right] \label{hm4}
\end{equation}
 for the vector field $\vec{n}(\vec r)= \{\sin\theta\cos\Phi ,\,\sin\theta\sin\Phi ,\,\cos\theta\}$,
obtained by Belavin and Polyakov \cite{BP} more than two decades ago. A
renewed interest to these unconventional spin textures is stimulated by
high-$T_c$ problem in doped quasi-2D-cuprates and  quantum Hall effect.

The skyrmion spin texture consists of a vortex-like arrangement of the in-plane
components of spin with the $z$-component reversed in the centre of the
skyrmion and gradually increasing to match the homogeneous background at
infinity. The spin distribution within classical skyrmion  is given as follows
\begin{equation}
\Phi=q\varphi\qquad\cos\theta=\frac{r^{2q}-\lambda^{2q}}{r^{2q}+\lambda^{2q}},
\label{sk}
\end{equation}
or for $q=1$
\begin{equation}
n_{x}=\frac{2r\lambda }{r^{2}+\lambda ^{2}}\cos \varphi ,\qquad
n_{y}=\frac{2r\lambda }{r^{2}+\lambda ^{2}}\sin \varphi ,\qquad
n_{z}=\frac{r^{2}-\lambda ^{2}}{r^{2}+\lambda ^{2}}.
\end{equation}
In terms of the stereographic variables the skyrmion with radius  $\lambda $   and
phase  $\varphi _{0}$  centered at a point $z_0$ is identified with spin distribution
 $w(z)=\frac{\Lambda }{z-z_0}$, where $z=x+iy=re^{i \varphi }$  is a point in the complex
 plane, $\Lambda =\lambda e^{i\theta}$,
and characterized by three modes: translational, or positional $z_0$-mode,
"rotational" $\theta$-mode and    "dilatational" $\lambda$-mode. Each of them
corresponds to a certain symmetry of the classical skyrmion configuration. For
example, $\theta$-mode corresponds to a combination of rotational symmetry and
internal U(1) transformation. Classical skyrmionic energy $E_q =8\pi qJS^2$ is
proportional to its topological charge and does not depend on its radius.

Other well known solutions of isotropic and anisotropic 2D Heisenberg model are
the in-plane and out-of-plane vortices \cite{Hb,Ivanov,Borisov} which have the
energy logarithmically dependent on the size of the system. The in-plane vortex
is described by the formulas $\Phi=q\varphi$, $\cos\theta=0$. The $\theta(r)$
dependence  for the out-of-plane vortex cannot be found analytically.

Non-linear $\sigma$-model can be addressed as classical continuum limit
 of  2D Heisenberg ferromagnet with isotropic  spin-Hamiltonian
\begin{equation}
H=-J\sum_{i,b}\hat{\vec{S}}_i \hat{\vec{S}}_{i+b}. \label{H}
\end{equation}

The simplest quantum generalization of skyrmionic solutions could be obtained
in frames of spin coherent states \cite{Perelomov},
where the wave function of the quantum spin system, which maximally corresponds to
classical skyrmion, is a product of spin coherent states. In the
case of spin $s=\frac{1}{2}$
\begin{equation}
\Psi _{sk}( 0) =\prod\limits_{i}\lbrack \cos \frac{\theta
_{i}}{2}e^{i\frac{\varphi _{i}}{2}}\mid \uparrow \rangle +\sin \frac{\theta
_{i}}{2}e^{-i\frac{\varphi _{i}}{2}}\mid \downarrow \rangle \rbrack ,
\label{CS}
\end{equation}
where $\theta_i=\arccos [(r_i^2-\lambda^2)/(r_i^2+\lambda^2)]$. The coherent
state implies a maximal equivalence to the classical one with the minimal
uncertainty of spin components. Actually, every on-site spin
 in a lattice is assumed to be subjected to a molecular field $\vec{H}(\vec r)\propto
 \vec{n}(\vec r)=
\{\sin\theta\cos\Phi ,\,\sin\theta\sin\Phi ,\,\cos\theta\}$  which spatial
distribution forms a skyrmionic texture.

The coherent state approach appears to be rather simple for the $s=1/2$ spin systems.
Indeed, on the one hand, spin-Hamiltonian for $s=1/2$ quantum system is restricted to have isotropic,
or anisotropic bilinear Heisenberg exchange form like (\ref{H}).  On the other
 hand, the trial wave function (\ref{CS}) is simply parameterized by the vector
field $\vec{n}(\vec r)$. Some quasiparticle properties of quantized skyrmion in
the $s=1/2$ model are addressed in \cite{Istomin}.

The situation becomes more involved for the $S\geq 1$ (pseudo)spin systems,
where, generally speaking, we have to deal with additional non-Heisenberg terms
in (pseudo)spin-Hamiltonian and several vector fields to parameterize the trial
wave function like (\ref{CS}). A principal difference between the
$S=\frac{1}{2}$ and $S\geq 1$ quantum systems lies in what concerns the order
parameters.
 The only single-site order parameter in the former case is
an average spin (dipole) moment $\langle S_{x,y,z}\rangle$, whereas in the
 latter one
 has  additional "spin-multipole" parameters like "spin-quadrupole" (spin nematic) averages
   $\langle \{\hat{S_i}\hat{S_j}\}\rangle$,
    where  $\{\hat{S_i}\hat{S_j}\}=\hat{S_i}\hat{S_j}+\hat{S_j}\hat{S_i}$. Hence, we may expect in
   $S=1$ quantum spin systems different
 topological defects with spatial distribution of not only nonzero
  spin (dipole) moment (dipole skyrmions), or spin-quadrupole moment (quadrupole
  skyrmions), but more involved dipole-quadrupole   skyrmions with a nonzero
 distribution of both order parameters.

    Interestingly, that in a sense,
  the  $s=1/2$ quantum spin system is closer to the classical one
    ($S\rightarrow \infty$) also characterized by the single-site vector order
     parameter than, for instance, the  $S=1$ quantum spin system with its eight site  order
parameters. In a whole, we should expect for the $S\geq 1$ (pseudo)spin
 systems an appearance both of unconventional
 topology and the complicated order parameter textures \cite{Vol}.

The main interest in $S=1$ quantum spin systems is provoked by the $S=1$
quantum spin chains displaying the Haldane gap \cite{Haldane}.  Several special
examples of the 2D $S=1$ spin systems have been extensively discussed earlier
in connection with the study of the static and dynamic properties of
anisotropic Heisenberg 2D magnets with a single-ion anisotropy (see {\it e.g.}
\cite{Loktev,Ivanov1}). Some aspects of the topological structure of vortices
in 2D $S=1$ systems were discussed    by different authors in connection with
$^3$He problem  \cite{Volovik,Rajantie,Vol} and triplet superconductivity
\cite{Knig}. The quantum "spin nematic" phase of the fully isotropic $S=1$
system was addressed recently by Ivanov and Kolezhuk \cite{nematic}.

Our interest in this field  was motivated by a problem of a description of
soliton-like excitations in quasi-2D cuprates in frames of a novel scenario
proposed by one of authors \cite{Moskvin}. The model considers doped cuprates
as a system of singlet local bosons moving in a lattice formed by hole centers
$CuO_{4}^{5-}$. Such a center has a complex ground state multiplet
$^{1}A_{1g}-{}^{1}E_{u}$ which can be described by a pseudo-spin $S=1$.

 In this
paper we make use of the spin coherent (SC-) state approach to describe the
skyrmion-like topological defects in $S=1$ quantum (pseudo)spin 2D systems with
isotropic non-Heisenberg spin-Hamiltonian.
 These solutions  are a special case of so-called CP$_N$
spinors   discussed in Refs.\cite{Raj,Raj2}, though its  authors focused their
interest  only in CP$_1$ and CP$_3$ models, rather than  the CP$_2$ spinors of
our model.
  Our interest is  mainly
focused on the unconventional quadrupole and dipole-quadrupole   skyrmion-like
static solutions.

%It should be noted that some special cases were addressed earlier in papers
%\cite{Knig}.  As a text-book example of the  topological defects in
%  spin-1 systems the vortices in $^3$He are best known \cite{Vol}.

In Sec.II we address the isotropic bilinear-biquadratic  Hamiltonian for the
$S=1$ quantum (pseudo)spin systems, the parameterization of the trial wave
function, the SU(3)-model approach, and the reduction procedure to the
classical continuum limit of the  $S=1$ model. In Sec.III the unconventional
skyrmion-like solutions are analyzed, including the known magnetic (dipole)
skyrmion \cite{Knig}, and unusual quadrupole and dipole-quadrupole skyrmions
with a non-trivial spatial distribution of dipole and/or quadrupole
(pseudo)spin order parameters.

\section{Classical description of the $S=1$ quantum (pseudo)spin systems}

In general, isotropic non-Heisenberg spin-Hamiltonian for the $S=1$ quantum
 (pseudo)spin systems should include both bilinear Heisenberg exchange term and
 biquadratic non-Heisenberg exchange term:
\begin{equation}
 \hat{H}=-\tilde J_1\sum_{i,\eta}\hat{\vec{S}}_i
\hat{\vec{S}}_{i+\eta}-\tilde
J_2\sum_{i,\eta}(\hat{\vec{S}}_i\hat{\vec{S}}_{i+\eta})^2= \label{ha1} 
\end{equation}
$$=-J_1\sum_{i,\eta}\hat{\vec{S}}_i
\hat{\vec{S}}_{i+\eta}-J_2\sum_{i,\eta}\sum_{k\geq
j}^3(\{\hat{S}_k\hat{S}_j\}_i\{\hat{S}_k\hat{S}_j\}_{i+\eta}) \label{ha1}
$$
where $J_i$ are the appropriate exchange integrals, $J_1=\tilde J_1-\tilde
J_2/2$, $J_2=\tilde J_2/2$, $i$ and $\eta$ denote the summation over lattice
sites and nearest neighbours, respectively. In our spin-1 model we use trial
functions 
\begin{equation}
 \psi=\prod_{j\in lattice}c_i(j)\psi_i=\prod_{j\in
lattice}(a_i(j)+ib_i(j))\psi_i \label{wf1} 
\end{equation} 
Here $j$ labels a lattice site
and the spin functions $\psi_i$ in cartesian
 basis are used: $\psi_z=|10>$ and $\psi_{x,y}\sim(|11>\pm|1-1>)/\sqrt 2$.
 The linear (dipole) spin-operator is represented by  simple matrix:
 $$
 <\psi_i|S_j|\psi_k> =-i\varepsilon_{ijk},
 $$
  and for the order parameters one
 easily obtains:
\begin{equation}
 <\hat{\vec S}> = -2[\vec a,\vec b], \qquad
<\{\hat{S_i}\hat{S_j}\}>=2(\delta_{ij}-a_ia_j-b_ib_j) \label{med} 
\end{equation}  
given the normalization constraint ${\vec a}^2 +{\vec b}^2=1$. Thus, for the case of
spin-1 system the order parameters are determined by two classical vectors (two
real components of one complex vector $\vec c =\vec a +i\vec b$ from
(\ref{wf1})). The two vectors are coupled, so the minimal number of dynamic
variables describing the $S=1$ spin system appears to be equal to four (see
Ref.\cite{Loktev} and Sec.2.2 below). Hereafter we would like to emphasize the
$director$ nature of the ${\vec c}$ vector field: $\psi ({\vec c})$ and $\psi
(-{\vec c})$ describe the physically identical states.

 The dipole or magnetic skyrmions in the spin-1 systems were
addressed in the paper \cite{Knig}.
   The structure of the order parameter admits the existence of more general
 types of solutions which are purely quadrupole ("electric") or mixed
 dipole-quadrupole ("magneto-electric") ones.
  One should note that, in common, the length of the spin vector in S=1 model must not
be fixed.
The order parameters structure is responsible for another important property
 of the S=1 systems: it allows the existence of more than one topological
 quantum number.

\subsection{SU(3)-symmetry model: Gell-Mann operators and effective Hamiltonian for S=1 model}

Three spin-linear (dipole) operators $\hat S_{1,2,3}$ and five independent
spin-quadrupole operators $\{{\hat S_{i}},{\hat S_{j}}\}-\frac{2}{3} {\hat
{\vec S}}^{2}\delta _{ij}$ at $S=1$ form eight Gell-Mann operators being the
generators of the SU(3) group. Below we will make use of the appropriate
Gell-Mann $3\times 3$
 matrices $\Lambda^{(k)}$,  which
differ from the conventional $\lambda^{(k)}$ only by a renumeration:
$\lambda^{(1)}=\Lambda^{(6)}$, $\lambda^{(2)}=\Lambda^{(3)}$,
$\lambda^{(3)}=\Lambda^{(8)}$, $\lambda^{(4)}=\Lambda^{(5)}$,
$\lambda^{(5)}=-\Lambda^{(2)}$, $\lambda^{(6)}=\Lambda^{(4)}$,
$\lambda^{(7)}=\Lambda^{(1)}$, $\lambda^{(8)}=\Lambda^{(7)}$.
First three matrices $\Lambda^{(1,2,3)}$ correspond to linear (dipole) spin
operators:
$$
\Lambda^{(1)}=S_x =
\pmatrix{0&0&0\cr
    0&0&-i\cr
    0&i&0\cr};
 \qquad  \Lambda^{(2)}=S_y=
\pmatrix{0&0&i\cr
    0&0&0\cr
    -i&0&0\cr};
    \qquad \Lambda^{(3)}=S_z=
\pmatrix{0&-i&0\cr
    i&0&0\cr
    0&0&0\cr};
$$
while other five matrices correspond to quadratic (quadrupole)  spin operators:
$$
   \Lambda^{(4)}=-\{S_zS_y\}=
\pmatrix{0&0&0\cr
    0&0&1\cr
    0&1&0\cr};
 \Lambda^{(5)}=-\{S_xS_z\}=
\pmatrix{0&0&1\cr
    0&0&0\cr
    1&0&0\cr};
 \Lambda^{(6)}=-\{S_xS_y\}=
\pmatrix{0&1&0\cr
    1&0&0\cr
    0&0&0\cr};
$$
$$
\Lambda^{(7)}=-\frac{1}{\sqrt{3}}(S_x^2+S_y^2-2S_z^2)=
\frac{1}{\sqrt{3}}\pmatrix{1&0&0\cr
    0&1&0\cr
    0&0&-2\cr};
\qquad \Lambda^{(8)}=S_y^2-S_x^2= \pmatrix{1&0&0\cr
    0&-1&0\cr
    0&0&0\cr};
$$
$$
S_x^2+S_y^2+S_z^2=2\hat E
$$

with $\hat E$ being a unit $3\times 3$ matrix.

The generalized spin-1 model can be described by the Hamiltonian  bilinear
 on the SU(3)-generators $\Lambda^{(k)}$
\cite{Sutherland}
$$
\hat{H}=-\sum_{i,i+\eta}\sum_{k=1}^{8}J_{km}\hat{\Lambda}_i^{(k)}
\hat{\Lambda}_{i+\eta}^{(m)} \, .
$$
Here $i,\eta$ denote lattice sites and nearest neighbors, respectively. This is
a $S=1$ counterpart of the $S=1/2$ model Heisenberg Hamiltonian with three
generators of  the SU(2) group or Pauli matrices included instead of eight
Gell-Mann matrices.
 %The
%spin 1/2 Hamiltonian commutes with the spin operator which is represented by
%the Pauli matrices being the . Spin is a conserving quantity. The higher
%degrees of spin operator are also conserved but all these can be expressed
%through the linear combinations of Pauli matrices.

In frames of isotropic bilinear-biquadratic  model we study in this work the
$8\times 8$ matrix $J_{km}$
 is assumed to be diagonal with
elements $J_{11}=J_{22}=J_{33}=J_1$, and $J_{44}=J_{55}=...=J_{88}=J_2$.
  Fully
isotropic SU(3) model with  $J_1 = J_2$ corresponds to a ferromagnetic version
of so-called Uimin-Lai-Sutherland model \cite{Sutherland}. It should be noted
that the isotropic in a real space Hamiltonian with $J_1\neq J_2$ can be
considered
 as the anisotropic one in the 8-dimensional SU(3) group space, and the symmetry of the model
 breaks to the subgroup SO(3)$\subset$SU(3). In other words, the breaking of the condition $J_1=J_2(=J)$ can
   be considered as an appearance of the exchange anisotropy in the 8-dimensional
   phase space.  To describe  this anisotropy one might
   introduce the ratio $\lambda =J_2 /J_1$. Hereafter, we shall associate the
   limiting   cases $\lambda =0$ and $\lambda \rightarrow \infty$ with purely magnetic
   and electric solutions, respectively. Such an effective anisotropy in $S=1$ systems
    differs strongly
   from the real spatial exchange anisotropy in $S=1/2$ systems which results
   in preferred spin orientations.

  If one considers the magnetic
 field parallel to z-axis or the anisotropy of exchange parameters in a real
  space the symmetry breaks to SO(2)$\subset$SO(3). However, given the definite
  relations between the anisotropy constants and exchange integrals this model
   can be reduced to spin-1/2 isotropic model. This case merits the separate
   examination.

\subsection{The classical continuum limit of S=1 model}

Having  substituted  our trial wave function (\ref{wf1}) to $\langle {\hat
H}\rangle$ provided $\langle \hat{\vec S}(1)\hat{\vec S}(2)\rangle =\langle
\hat{\vec S}(1)\rangle \langle \hat{\vec S}(2)\rangle $ we arrive at
 the Hamiltonian of the isotropic classical spin-1 model in the continuum
  approximation as follows:
\beq H=J_1\int d^2\vec r\left[\sum_{i=1}^{3}(\vec{\nabla}S_i)^2\right]+J_2\int
d^2\vec r\left[\sum_{i,j=1}^{3}(\vec{\nabla}a_ia_j+\vec{\nabla}b_ib_j)^2\right]
+\frac{4(J_2-J_1)}{c^2}\int |\vec{S}|^2d^2\vec r \, , \label{ham7} \eeq where
$\vec S=-2[\vec a*\vec b]=\langle \hat{\vec S}\rangle$. The  effective exchange
anisotropy defines the third "gradientless" term in the Hamiltonian that breaks
the scaling   invariance of the model. Such an effect in $S=1/2$ system appears
due to the real spatial exchange anisotropy  which defines the magnetic length
\cite{Borisov}.

The spin-1 model differs from the spin-1/2 model due to the appearance of the
additional ("nonmagnetic") degrees of freedom. When $\langle \hat{\vec
S}\rangle =0$ (e.g., if one of $\vec a,\vec b$ vectors turns into zero) we have
a non-zero part of classical Hamiltonian (proportional to $J_2$) and can get
the nontrivial configurations
 of non-zero vector. We shall call this configuration "electric skyrmion".
 It should be described by one vector with the fixed length, so the
 topological classification of such solutions is completely analogous to that
  of the spin-1/2 classical solutions, which are described by the order
  parameter being a fixed-length spin vector. Below, we shall study
   this solution, deriving it via reducing our biquadratic model to the
   non-linear O(3)-model. The topological charge of the classical electric
    skyrmion with $\vec b=0$ can be defined by  usual formula as follows
\beq Q=\frac{1}{8\pi}\int d^2\vec r\varepsilon_{\nu\mu}(\vec a*[\vec
a_{\nu}*\vec a_{\mu}]) =\frac{1}{4\pi}\int rdrd\varphi(\vec a*[\vec
a_{r}*\frac{1}{r}\vec a_{\varphi}])\, ,
 \label{chrg} \eeq
 where the subscripts denote derivative.

In the continuum limit for $J_1=J_2=J$ the Hamiltonian (\ref{ham7}) can be
transformed into the classical Hamiltonian of the $SU(3)$-symmetric
scale-invariant  model:
$$
H=\frac{1}{2}J\int d^2\vec r \,
[\sum_{k=1}^8(\vec{\nabla}\bar{\psi}\hat{\Lambda}^{(k)}\psi)^2] \, .
$$
where we make use of  the single-site wave function in the form as follows:
\beq \psi=\pmatrix{R_1\exp(i\Phi_1)\cr R_2\exp(i\Phi_2)\cr
R_3\exp(i\Phi_3)}\,;\qquad |\vec R|^2=1\, , \label{fun}
 \eeq
 with $\vec
R=\{\sin\Theta\cos\eta ,\sin\Theta\sin\eta ,\cos\Theta\}$. In accordance with
the $director$ nature of the ${\vec a}, {\vec b}$ vector fields we have to vary
the angles $\Theta ,\eta ,\Phi_i$ in the range $(0, \pi)$.  The Hamiltonian can
be rewritten as follows:
$$
H_{isotr}=2J\int d^2\vec r \{
(\vec{\nabla}\Theta)^2+\sin^2\Theta(\vec{\nabla}\eta)^2+
$$
\beq +\sin^2\Theta\cos^2\Theta\left[
\cos^2\eta(\vec{\nabla}\Psi_1)^2+\sin^2\eta(\vec{\nabla}\Psi_2)^2\right]
+\sin^4\Theta\cos^2\eta\sin^2\eta(\vec{\nabla}\Psi_1-\vec{\nabla}\Psi_2)^{2}\}\,
, \label{iso}
 \eeq
 where we have introduced $\Psi_1=\Phi_1-\Phi_3,\Psi_2=\Phi_3-\Phi_2$. For an
 isolated (pseudo)spin system the $\Phi_3$  phase is arbitrary,
  hence the minimal number of dynamic variables describing the $S=1$ (pseudo)spin system
  equals to four (=$4S$ \cite{Loktev}). However, for a more general situation, when
  the (pseudo)spin system represents only the part of the bigger system, and we are
  forced to consider the coupling with the additional degrees of freedom,  the $\Phi_3$  phase
 turns into a non-trivial parameter: $\Phi_3=Q_3\varphi$.
 The topological solutions for our Hamiltonian (\ref{iso}) can be classified  at least by
three topological quantum numbers (winding numbers): phases $\eta,\Psi_{1,2}$
can change
 by $2\pi$ after the passing around the center of the defect. It should be noted that
 the $director$ nature of the ${\vec a}, {\vec b}$ vector fields implies
 the possibility  for    winding numbers to take  half-integer  values.
 The appropriate  modes may have very complicated topological structure due
 to the possibility for  one defect to have several different centers
 (while one of the phases $\eta,\Psi_{1,2,3}$ changes by $2\pi$ given
  one turnover
 around one center $(r_1,\varphi_1)$, other phases may pass around other
 centers $(r_i,\varphi_i)$).
  Each center in a multi-center defect can be considered as a
 quasiparticle. In this connection it should be noted that the spin-1 model
  differs from the spin-1/2 one by the fact
 that  the former in common assumes quasiparticles of different types due
 to the  existence of different topological quantum numbers for different centers.
% We shall not analytically investigate such multi-center complicated defects
% in this work.

Finally, it should be noted that the above model approach can be extended to the
 $S>1$ isotropic (pseudo)spin systems.
 In the case of spin S one
 has to calculate averages like  $<S_{i_1}..S_{i_n}>$ where $n=1,..,2S$.
 One can suggest that in our Hamiltonian the combinations $c_{i_1}..c_{i_n}$
 with a fixed-length vector $\vec c$ should appear.
Now let us notice that for any $k\in N$ the model with discrete Hamiltonian
\beq
H_{kk}=-J_{kk}\sum_{i,\eta}\sum_{j_1..j_k=1}^{3}\left[\prod_{q=1}^{k}c_{j_q}(i)c_{j_q}(i+\eta)\right]
\label{hm2}
\eeq
in the continuum limit can be reduced to
\beq
H_k=J_k\int d^2\vec r\left[\sum_{j_1..j_k=1}^{3}(\vec{\nabla}c_{j_1}..c_{j_k})^2\right]\, .
\label{hm3}
\eeq
Making use of $\vec{\nabla}(fg)=f\vec{\nabla}g+g\vec{\nabla}f$ provided
 $|\vec c|=$const, we come to
\beq H_k=kJ_k|\vec c|^{2k-2}\int d^2\vec
r\left[\sum_{i=1}^{3}(\vec{\nabla}c_i)^2\right]\, . \label{hm4}
 \eeq
  It is a
non-linear O(3)-model. Its skyrmionic solutions differ from the conventional
 ones by the energy, due to the term $k|\vec c|^{2k-2}$. In our spin-1
 case $k=2$, $\vec c=\vec n$. So, if the reduction of spin-S quantum model
  to the quasi-classical one (\ref{hm2}) will be made by means of some
  parameterization
of the trial wavefunction it will be easy to obtain the skyrmionic
solutions of the model with finite energy $kJ_kq|\vec c|^{2k}$.

\section{Unconventional skyrmions in $S=1$ (pseudo)spin systems}

\subsection{Dipole skyrmions}

One important case of the spin-1 model when $J_2=0$ (purely Heisenberg
Hamiltonian) also has skyrmionic solutions, which were found earlier in Ref.
\cite{Knig}.  When  the $\vec a,\vec b$ vectors are perpendicular to each other
($\vec a\perp \vec b$), the model also reduces to the nonlinear
 O(3)-model. The solution from Ref. \cite{Knig} is described by the following formulas
 (in polar coordinates):
$$
\sqrt{2}\vec a=(\vec e_z\sin\theta-\vec e_{r}\cos\theta)\sin\varphi+
\vec e_{\varphi}\cos\varphi \, ;
$$
\beq
\sqrt{2}\vec b=(\vec e_z\sin\theta-\vec e_{r}\cos\theta)\cos\varphi-
\vec e_{\varphi}\sin\varphi \,.
\label{knig}
\eeq
 The fixed-length spin vector  is distributed in the same way as in
the conventional skyrmion with topological charge $q=1$ (\ref{sk}). However,
unlike the usual skyrmions, the solutions
 (\ref{knig}) have additional topological structure due to
 the existence of two vectors $\vec a$ and $\vec b$. Going around the center
 of the defect the vectors $\vec a$ and $\vec b$ can make $N$ turns around the
 spin vector $\propto [{\vec a}\times{\vec b}]$. Thus, we can introduce two topological
 quantum numbers: $N$ and $q$ \cite{Knig}. In addition, it should be noted that
 $q$  number  may be half-integer.
 Hereafter we shall call the skyrmionic solutions with the only non-zero magnetic
 component as the dipole or magnetic ones.

\subsection{Quadrupole skyrmions}

 Magnetic skyrmions as the solutions of purely Heisenberg
 (pseudo)spin Hamiltonian \cite{Knig} were obtained given the restriction
   $\vec a\perp\vec b$ and the lengths of these vectors were fixed.
   These restrictions lead to the pure magnetic solution and enabled
    to use a subgroup for the topological
    classification \cite{Knig}. It was SO(3) which
 is generated by  $\hat{\Lambda}^{(1)},\hat{\Lambda}^{(2)},
 \hat{\Lambda}^{(3)}$ matrices forming the Heisenberg bilinear term.

   Hereafter we address another situation with purely biquadratic
  (pseudo)spin Hamiltonian ($J_1$=0) and treat the
    non-magnetic (electric) degrees of freedom.
    The topological classification of the purely electric
    solutions is simple because it is also based on the making use  of subgroup
     instead of the full group. We address the solutions given
      $\vec a\parallel\vec b$ and the fixed
      lengths of the vectors, so we use for the
     classification the same subgroup as above.

  The biquadratic part of the Hamiltonian
\beq H_{biq}=J_2\int d^2\vec
r\left[\sum_{i,j=1}^{3}(\vec{\nabla}a_ia_j+\vec{\nabla}b_ib_j)^2\right]
\label{hamEl} \eeq can be rewritten as follows \beq H_{biq}=J_2\int d^2\vec
r\left[ \sum_{i,j=1}^{3}(\vec{\nabla}n_in_j)^2\right]\,, \label{hm1} \eeq where
$\vec a=\alpha\vec n, \vec b=\beta\vec n$, and $\alpha+i\beta= \exp (i\kappa)$,
$\kappa\in R$. We denote $\vec n=n\{\sin\Theta\cos\Phi, \, \sin\Theta\sin\Phi,
\, \cos\Theta\}$. Using  simple formula
$\vec{\nabla}(fg)=f\vec{\nabla}g+g\vec{\nabla}f$ together with the
normalization constraint $|\vec n|^2=$const, we reduce the expression for
$H_{biq}$ to the familiar nonlinear O(3)-model: \beq H_{biq}= 2J_2|\vec
n|^{2}\int d^2\vec r\left[\sum_{i=1}^{3}(\vec{\nabla}n_i)^2\right]\, .
\label{hm4} \eeq
 Its solutions are skyrmions, but instead of
the spin distribution in magnetic skyrmion we have  solutions with zero spin,
 but the non-zero distribution of five spin-quadrupole moments
  $\langle \Lambda ^{(4,5,6,7,8)}\rangle$, or
   $\langle \{S_{i}S_{j}\}\rangle$ which in turn are determined
by the distribution of the $\vec a$($\vec n$) vector:
\beq
 \Phi=q\varphi+\Phi_0 ; \quad \cos\Theta=\frac{r^{2q}-\lambda^{2q}}{r^{2q}+\lambda^{2q}}
\label{slnEl} \eeq with a classical skyrmion energy \beq E_{el}=16\pi qJ_{2}\,
. \eeq The distribution of the spin-quadrupole moments $\langle
\{S_{i}S_{j}\}\rangle$
 can be easily obtained:
$$
\langle S_x^2\rangle =\frac{(r^{2q}+\lambda^{2q})^2\cos^2q\varphi+(r^{2q}-\lambda^{2q})^2\sin^2q\varphi}{(r^{2q}+\lambda^{2q})^2}\,;
$$
$$
\langle S_y^2\rangle =\frac{(r^{2q}+\lambda^{2q})^2\sin^2q\varphi+(r^{2q}-\lambda^{2q})^2\cos^2q\varphi}{(r^{2q}+\lambda^{2q})^2}\,;
$$
$$
\langle \{S_xS_y\}\rangle =-\frac{2r^{2q}\lambda^{2q}\sin2q\varphi}{(r^{2q}+\lambda^{2q})^2}\,;
\qquad
\langle S_z^2\rangle =\frac{4r^{2q}\lambda^{2q}}{(r^{2q}+\lambda^{2q})^2}\,;
$$
\beq \langle \{S_xS_z\}\rangle =\frac{2(\lambda^{2q}-r^{2q})r^q\lambda^q\sin
q\varphi}{(r^{2q}+\lambda^{2q})^2}\,; \qquad \langle \{S_yS_z\}\rangle
=\frac{2(\lambda^{2q}-r^{2q})r^q\lambda^q\cos
q\varphi}{(r^{2q}+\lambda^{2q})^2}\,.
\label{qq}
 \eeq
  One should be emphasized
that the distribution of five independent quadrupole order parameters for the
electric skyrmion are straightforwardly determined by a single vector field
${\vec n}({\vec r})$. The phase factor $\alpha+i\beta$ can be arbitrary,
because it is not included in the Hamiltonian. It may be written as follows:
$\alpha+i\beta=\exp(ip\varphi)$. Such classes of
 solutions are physically equivalent in frames of our model.

\subsection{Dipole-quadrupole skyrmions}

In this subsection we would like to write out some solutions of fully
SU(3)-symmetric
isotropic model for (pseudo)spin Hamiltonian (\ref{ham7}) with $J_1 =J_2$.
 We do not derive all the possible solutions of this model
 because  it merits  to be a subject of the separate examination. Our main goal
  now is only to illustrate the richness of the model using as examples the
   simplest solutions. The solutions of this model may be classified taking
    into account whether they have each of the winding numbers to be zero or
    not. Below we will briefly consider two simplest classes of such
     solutions (the choice of classes is defined by the simplicity of
     integration).

First of all we shall consider simplest solutions of the equations minimizing
energy functional. One type of skyrmions  can be obtained given the trivial
phases $\Psi_{1,2}$. If these are constant, the $\vec R$ vector distribution
(see ($\ref{fun}$)) represents the skyrmion described by the usual formula
(\ref{sk}).
  All but one topological quantum numbers are zero for this class of solutions.
  It includes both dipole and quadrupole solutions: depending
  on selected constant phases one can obtain both "electric" and different
  "magnetic" skyrmions. The substitution $\Phi_1=\Phi_2=\Phi_3$ leads to
  the electric skyrmion which was obtained as a solution of more general
   SU(3)-anisotropic model in the previous section. Another
   example can be $\Phi_1=\Phi_2=0,\Phi_3=\pi/2$. This substitution
   implies  $\vec b\Vert Oz,\vec a\Vert Oxy,\vec S\Vert Oxy$, and
   ${\vec S}=\sin\Theta\cos\Theta \{\sin\eta,-\cos\eta,0\}$.
   Nominally, this is the in-plane spin vortex with a varying length of
   the spin vector
$$
|{\vec S}|=\frac{2r\lambda|r^2-\lambda^2|}{(r^2+\lambda^2)^2}\,,
$$
which turns into zero at the circle $r=\lambda$, at the center $r=0$ and at
 the infinity $r\rightarrow \infty$,
and has maxima at  $r=\lambda(\sqrt 2 \pm 1)$. In addition to the non-zero
in-plane components of spin-dipole moment $\langle S_{x,y}\rangle$ (or $\langle
\Lambda^{(1,2)}\rangle$) this vortex is characterized by a non-zero
distribution of (pseudo)spin-quadrupole moments $\langle
\Lambda^{(6,7,8)}\rangle$.

Here we would like to emphasize the difference between spin-1/2 systems in
 which there are such the solutions as in-plane vortices with the energy
  having a well-known logarithmic dependence on the size of the system
   and fixed spin length, and spin-1  systems in which the in-plane vortices
    also can exist but they may have a finite energy and a varying spin length.
     The distribution of quadrupole components associated with in-plane
      spin-1 vortex is non-trivial. Such solutions can be named as
      "in-plane dipole-quadrupole skyrmions".

Other type of solutions we get given the phases
$\Psi_1=Q_1\varphi,\Psi_2=Q_2\varphi$ with two integer winding numbers
$Q_{1,2}$ and $\eta=\eta(r), \Theta=\Theta(r)$. Easiest way to obtain the
solutions of variational equations is to put one of these functions to be
constant. This way we arrive at six types of solutions. The first one
$\Theta=0$ corresponds to the trivial spinor \beq
(i):\,\psi=\pmatrix{R_1\exp(i\Phi_1)\cr R_2\exp(i\Phi_2)\cr R_3}=\pmatrix{0\cr
0\cr 1}\, .
 \label{1}
 \eeq
 Other solutions being written in complex form with $z=re^{i\varphi}/\lambda$
 ($\bar{z}=z^*$)
are as follows:
 \beq
(ii):\,\Theta=\pi/2+\pi k, k\in Z, \eta(z)=\arctan [|z|^{(q_1+q_2)}];\,
 \psi=\frac{exp(iq_1\varphi)}{(1+|z|^{2(q_1+q_2)})^{1/2}}\pmatrix{1\cr (-1)^k\bar{z}^{(q_1+q_2)}\cr
0}; \label{2}
 \eeq
\beq (iii):\,\eta=\pi k, k\in Z,\Theta(z)=\arctan [|z|^{q_1}];\,
 \psi=\frac{1}{(1+|z|^{2q_1})^{1/2}}\pmatrix{(-1)^k\bar{z}^{q_1}\cr0 \cr
1}; \label{3}
 \eeq
\beq (iv):\,\eta=\pi/2+\pi k, k\in Z,\Theta(z)=\arctan [|z|^{q_2}];\,
 \psi=\frac{1}{(1+|z|^{2q_1})^{1/2}}\pmatrix{0\cr(-1)^k\bar{z}^{q_2}\cr
1}; \label{4}
 \eeq
\beq (v):\,\eta=\pi/4+\pi k/2, k\in Z,q_1=-q_2=q,\Theta(z)=\arctan [|z|^{q}];\,
 \psi=\frac{1}{(1+|z|^{2q})^{1/2}}\pmatrix{z^q/\sqrt{2}\cr z^q/\sqrt{2}\cr
1}; \label{5}
 \eeq
\beq (vi):\,\eta=\pi/4+\pi k/2, k\in Z,q_1=q_2=q,\Theta(z)=\arccos
[\frac{|z|^{2q}-1}{|z|^{2q}+1}];\,
 \psi=\frac{1}{(1+|z|^{2q})}\pmatrix{z^q\sqrt{2}\cr \bar{z}^q\sqrt{2}\cr
1-|z|^{2q}}; \label{6}
 \eeq

The energy for solution $(vi)$ is  $E_6=16\pi qJ$, while for others
($(ii)-(v)$)
 $E_k=4\alpha_k\pi^2J$ with $\alpha_2=q_1+q_2$, $\alpha_3=q_1$,
$\alpha_4=q_2$ and  $\alpha_5=q$. Similar solutions,  as so-called CP$_N$
spinors
 \beq
 \eta (z)=\frac{1}{(a^2+Nr^{2q})^{1/2}}\pmatrix{a\cr z^q\cr
z^q\cr \cdot\cdot\cdot\cr \cdot\cdot\cdot\cr z^q}\,, \label{eta}
 \eeq
($q$ is a winding number, $a$ is any constant) are discussed in
Ref.\cite{Raj,Raj2}, though its authors focused their interest
 only in CP$_1$ and CP$_3$ models, rather than  the CP$_2$ spinors of our model.

Above we address the dipole-quadrupole solutions with the only angular
parameter having the $z(r,\varphi)$-dependence. Hereafter, we consider the
nontrivial solution with two winding numbers and unconventional $z$-dependence
of angular parameters. To this end we make use of a $\psi$-spinor in another
form as follows:
\begin{equation}
\psi \,=\,
\pmatrix{\sin(\frac{\theta}{2})\exp(i\frac{\alpha}{2})(\cos(\frac{\beta}{2})\sin(\frac{\Phi}{2})
+i\sin(\frac{\beta}{2})\cos(\frac{\Phi}{2}))\cr
\sin(\frac{\theta}{2})\exp(i\frac{\alpha}{2})(\sin(\frac{\beta}{2})\sin(\frac{\Phi}{2})
-i\cos(\frac{\beta}{2})\cos(\frac{\Phi}{2}))\cr \cos(\frac{\theta}{2})\cr}.
\end{equation} In the system where $\hat S_z$ is diagonal it takes form
$$
\pmatrix{\sin(\frac{\theta}{2})\cos(\frac{\Phi}{2})\exp(i\frac{\alpha+\beta}{2})\cr
\sin(\frac{\theta}{2})\sin(\frac{\Phi}{2})\exp(i\frac{\alpha-\beta}{2})\cr
\cos(\frac{\theta}{2})\cr}.
$$
Such a parametrization is opportune when we are interested in solutions with
mean-values of $S_z$ and $Q_{zz}$ independent of $\varphi$. The density of the
energy functional with $J_1=J_2=1$ can be written out as follows:
$$
W=\frac{1}{2}[(\vec{\nabla}\theta)^2+\frac{\sin^2\theta}{4}\{(\vec{\nabla}\alpha)^2
+2\sin\Phi(\vec{\nabla}\alpha,\vec{\nabla}\beta)+(\vec{\nabla}\beta)^2\}+
$$
\begin{equation}
+\sin^2\frac{\theta}{2}\{(\vec{\nabla}\Phi)^2+
\cos^2\Phi\sin^2\frac{\theta}{2}(\vec{\nabla}\beta)^2\}].
\end{equation}
The substitution into the equations minimizing this functional of
$\alpha=q\varphi$, $\beta=p\varphi$ and $\theta=\theta(r),\Phi=\Phi(r)$
immediately satisfies two of them; the remaining two are as follows:
$$
4(\theta_{rr}+\frac{\theta_r}{r})=\sin\theta\Phi_r^2+\frac{\sin\theta}{r^2}
[p^2\cos^2\Phi+\cos\theta(q+p\sin\Phi)^2];
$$
\begin{equation}
\Phi_{rr}+\frac{\Phi_r}{r}=-\mbox{ctg}\frac{\theta}{2}\theta_r\Phi_r
+\frac{p\cos\Phi}{r^2}(q\cos^2\frac{\theta}{2}-p\sin^2\frac{\theta}{2}\sin\Phi)
\, .
\end{equation}
The numbers $\frac{p+q}{2}$ and $\frac{p-q}{2}$ are to be clearly integer. The
energy takes the following form:
\begin{equation}
W=\theta_r^2+\sin^2\frac{\theta}{2}\Phi_r^2+\frac{\sin^2\theta}{4r^2}[q^2+2pq\sin\Phi+p^2]+
\frac{p^2}{r^2}\sin^4\frac{\theta}{2}\cos^2\Phi \, .
\end{equation}
To find solutions with the finite energy one must choose the conditions
$\theta\to k\pi$ at the  infinity with k being integer. The second equation
implies that at the infinity $\Phi\to (2m+1)\pi/2$ when $\theta\to k\pi$. For
the simplest case with minimal different values of the winding numbers:
$p=1,q=3$ we have found the exact solution
\begin{equation}
\sin\Phi =\pm \frac{\lambda^2-r^2}{\lambda^2+r^2}; \qquad \cos\theta(r)=\pm
\frac{r^4(l^2+\lambda^2)-(r^2+\lambda^2)l^4}{r^4(l^2+\lambda^2)+(r^2+\lambda^2)l^4}.
\end{equation}
The energy of the solution does not depend on two radii $\lambda$ and $l$ and
equals to $16\pi$ (half the energy of the electrical skyrmion). The asymptotic
behaviour of the spinor is (two cases correspond to two signs for $\cos\theta$)
at zero and at the infinity
$$
\mbox{from}\pmatrix{0\cr 0\cr
1\cr}\qquad\mbox{to}\qquad\pmatrix{\frac{\exp(2i\varphi)}{\sqrt 2 }\cr
\frac{\exp(i\varphi)}{\sqrt 2}\cr 0\cr},\mbox{or}\,\,
\mbox{from}\pmatrix{\frac{\exp(2i\varphi)}{\sqrt 2 }\cr
\frac{\exp(i\varphi)}{\sqrt 2}\cr 0\cr}\qquad\mbox{to}\qquad\pmatrix{0\cr 0\cr
1\cr},
$$
respectively.

Now let us discuss the solutions from other point of view. In SU(3)-isotropic
case the density of energy is invariant under the transformations
$$
 \hat T=\exp[i\sum_{k=1}^{8}\gamma
_k\hat{\Lambda}^{(k)}]\, ,
$$
 generated by $\{\Lambda^{(k)}\}$, where eight parameters $\gamma _k$ do not depend on the coordinates.
If $\psi_2=\hat T\psi_1$, then
$$
\sum_{k=1}^{8} (\psi_1\hat{\Lambda}^{(k)}\psi_1)^2= \sum_{k=1}^{8}
(\psi_2\hat{\Lambda}^{(k)}\psi_2)^2
$$
at each point. In other words, both $\psi_1$ and $\psi_2$ are the solutions of
variational equations  with the same energy. Hence, all the solutions can be
classified on the irreducible representations of SU(3) group, or
supermultiplets. Generally speaking, two spinors $\psi_1$ and $\psi_2$ from the
same supermultiplet  represent only one solution in different
parameterizations. For instance, five solutions $(ii) - (v)$ represent one
solution $\{\psi_1,\psi_2,0\}$ which is in fact a well known Belavin-Polyakov
solution constructed on two components.
%It is also correct for CP$_N$ solutions
%mentioned by Rajaraman \cite{Raj}. In adjoint representation all these
%solutions are eight component qualities. For brevity we refer to this quality
%as 8-vector (strictly speaking it is not a vector). The components of this
%8-vector are 3 spin-dipole and 5 spin-quadrupole mean values. Among the
%functions representing the spatial dependence of these components there are
%three linearly-independent ones:
%$\sin\theta\cos\varphi,\sin\theta\sin\varphi,\cos\theta$
% with $\cos\theta=(r^{2}-\lambda^2)/(r^{2}+\lambda^2)$. Due to this fact the
%8-vector can be after appropriate transformation reduced to the 8-vector with 3
%non-zero spatially-dependent (with the same dependence as the spin-components
%of SU(2)-skyrmion), one constant ($1/\sqrt 3$) and 4 zero components. Non-zero
%components should be necessarily mixed dipole-quadrupole ones. The quadrupole
%part of the 8-vector is always non-zero, because the length of the classical
%spin vector cannot be more than 1, and the length of the full 8-vector is 4/3.
%The magnetic component should be present in this solution because three
%spatially-dependent components are constructed on three operators which
%generate SU(2) subgroup of SU(3). In trivial solution there could be six zero
%components, 1 and $1/\sqrt 3$.
% To construct
%a solution of this model from SU(2) skyrmion one should add to this spinor the
%third component (zero) and to make an 8-parametric transformation $\hat T$. All
%the solutions constructed from SU(2) spinor correspond to one class.

The solutions like electric skyrmion (see Eqs.(\ref{slnEl})-(\ref{qq})) and
solution $(vi)$ from this section correspond to another supermultiplet and can
be transformed into each other by a rotation around $z$-axis. Solutions with
the $\vec R$ vector distribution (\ref{1}) derived in the beginning of this
section can be reduced to electric skyrmion by the transformations generated by
$\hat{\Lambda}^{(4)}$ and $\hat{\Lambda}^{(8)}$.
%These solutions cannot be reduced to the "SU(2)"-solutions, because the number
%of linearly-independent functions is different. The phases of spinor
%$\psi_{SU(2)}$-components are $q\varphi$ and constant. The phases of $\psi_6$
%are $q\varphi$,  $-q\varphi$ and constant.
Another non-trivial supermultiplet  can be constructed from magnetic skyrmion
(\ref{knig}). More detailed analysis of the SU(3) supermultiplets will be given
elsewhere.

 Concluding this subsection we have to notice  that the
dipole-quadrupole solutions can be obtained also in more general $J_1\neq J_2$
model but the corresponding equations due to the term proportional to $J_2-J_1$
cannot be solved analytically similarly the  situation  in S=1/2 model with
exchange anisotropy for the out-of-plane vortices \cite{Hb}. It should be noted
that varying the "anisotropy" parameter $\lambda =J_2 /J_1$ from $\lambda =0$
to $\lambda \rightarrow\infty$ we deal with a  transformation of purely
magnetic solution to a purely electric one. One might expect that magnetic
skyrmion (\ref{knig}) would be stable given $0\leq\lambda <\lambda _{c_1}$,
while electric one would be stable given $\lambda _{c_2}\leq\lambda <\infty$.
In the intermediate range $\lambda _{c_1}\leq\lambda <1$ and $1<\lambda
<\lambda _{c_2}$ we expect the stability of dipole-quadrupole skyrmions with
predominant "magnetic" and "electric" behavior at infinity, respectively. Given
$\lambda =1$ we come to the fully SU(3) isotropic solution. The critical values
$\lambda _{c_{1,2}}$ of the "anisotropy" parameter should be calculated
numerically similarly to S=1/2 anisotropic Heisenberg model \cite{Gouvea}.
%There exist at least 3 non-trivial classes of the solutions of SU(3)-isotropic
%model.

\section{Conclusion}

In terms of spin coherent states  we have investigated topological defects
 in 2D  $S=1$ (pseudo)spin quantum
system
with the bilinear and biquadratic isotropic exchange in the continuum limit.
The proper Hamiltonian of the model can be written as bilinear on
 the generators of SU(3) group (Gell-Mann matrices). Knowledge of such group structure enables us to
 obtain some new exact analytical results.
The analysis of the proper classical model and its topology allows us to get
different skyrmionic solutions with finite energy and the spatial distribution
of spin-dipole and/or spin-quadrupole moments termed as dipole, quadrupole, and
dipole-quadrupole skyrmions, respectively. Among the latter  we would like to
note the in-plane  vortices with the in-plane distribution of spin moment,
varying spin length, and the non-trivial distribution of spin-quadrupole
moments.

One should note that for traditional spin systems like $3d$ magnetic oxides
with $S\geq 1$ the biquadratic exchange as a rule two orders of magnitude
smaller as compared with usual Heisenberg bilinear isotropic exchange. It
seems, for such systems the above analysis may play only purely theoretical
interest. However, for systems with the non-quenched orbital moments,
(pseudo)-Jahn-Teller effect, and other forms of (pseudo)degeneracy the
effective pseudo-spin Hamiltonian given $S\geq 1$ may include different large
non-Heisenberg terms. Namely for such systems we may expect the manifestation
of  unusual topological defects, including those addressed above.

\section{Acknowledgments}
We thank  A.B. Borisov, B.A. Ivanov, and A.P. Protogenov for valuable
discussions. The authors acknowledge a partial support from the INTAS Grant No.
01-0654, CRDF Grant No. REC-005, Russian Ministry of Education, Grant No.
E00-3.4-280 and No. UR.01.01.042, and Russian Foundation for Basic Researches,
Grant No. 01-02-96404. One of us (A. M.) would like to thank the Institut
f\"{u}r Festk\"{o}rper- und Werkstofforschung Dresden for hospitality.

\end{document}